# Bipublishers
## Bibliometric Indicators for Publishers

# Data processing, indicators and interpretation

Here we describe the Bibliometric Indicators for Publishers Project, an initiative undertaken by EC3Metrics SL for the analysis and development of indicators based on books and book chapters. Its goal is to study and analyze the publication and citation patterns of books and book chapters considering academic publishers as the unit of analysis. It aims at developing new methodologies and indicators that can better capture and define the research impact of publishers. It is an on-going project in which data sources and indicators are tested. We consider academic publishers as an analogy of journals, focusing on them as the unit of analysis. In this working paper we present the http://bipublishers.es/ website where all findings derived from the project are displayed. We describe the data retrieval and normalization process and we show the main results. A total 482,470 records have been retrieved and processed, identifying 342 publishers from which 254 have been analyzed. Then six indicators have been calculated for each publisher for four fields and 38 disciplines and displayed.

## Authors:

Nicolás Robinson-García

Evaristo Jiménez-Contreras

Enrique Fuente-Gutiérrez

Daniel Torres-Salinas[1]

[1] torressalinas@gmail.com

# Content

# 1. Introduction

Monographs and book chapters are key publication types in many fields of the Social Sciences and Arts & Humanities. Their use as communication channels is also common in other scientific fields, despite being journal articles their main publication type (Milojevic et al., 2014). Until recently, bibliometricians have excluded them from their analyses mainly due to the lack of bibliometric data sources containing monographs. As a consequence, publication and citation patterns in monographs have not been as studied as in journals. There is consensus on the fact that current bibliometric indicators do not apply as well in this context as they do with journal articles. Also, the fact that they are more expanded in the Social Sciences and the Arts & Humanities, emphasizes differences between these fields and the rest.

The 'Bibliometric Indicators for Publishers' project (hereafter BiP) is an initiative aimed at developing new methodologies and indicators that can better capture and define the research impact of books and book chapters. It is an on-going initiative in which data sources and indicators are tested. We consider academic publishers as an analogy of journals, focusing on them as the unit of analysis; an approach already suggested elsewhere (i.e., Giménez-Toledo & Román-Román, 2009; Torres-Salinas & Moed, 2009). We include six indicators for more than 100 publishers in four broad fields and 38 different disciplines. The data is based on the Thomson Reuters' Book Citation Index.

The Book Citation Index (hereafter BKCI) was released in 2011 aiming to shred light on the research performance of monographs. It filled a gap which was already noted by Garfield (1996), creator of the original Science Citation Index. Since its launch, several studies have been conducted describing and analyzing the strengths and caveats of this unique bibliometric database (Gorraiz, Purnell & Glänzel, 2013; Leydesdorff & Felt, 2012; Torres-Salinas et al., 2012;2013;2014a;2014b). These studies have reported strong limitations regarding its coverage, concentration of publishers and language biases. The findings reported in these studies should be taken into account when analyzing the results of this project.

Most analyses on citation and publication patterns have focused on specific disciplines comparing between journals and monographs and emphasizing differences between these disciplines. This project does not aim to rank publishers according to a given performance indicator, but to develop different types of indicators that may capture different characteristics of academic publishers. Here we offer a description of the results of the BiP project.

This paper is structured as follows. Section 2 describes the functionalities and options displayed in the web platform available at http://bipublishers.es. It also details its potential interest and the targeted audience to which it could be of interest. Section 3 describes the BKCI, the construction of fields and disciplines, the normalization process followed to identify publishers, and a definition of the indicators provided. In Section 4 we offer a global overview of the data analyzed. We display results for four fields (Science, Engineering & Technology, Social Sciences and Humanities & Arts), by indicator and by publishers. Finally, we briefly discuss the most important highlights in section 5. Aditionally, we have included Appendix A with detailed information on the aggregation of subject categories for the construction of each field and discipline.

## 2. Web platform and target audience

All results derived from the BiP project are made available through its website at http://bipublishers.es. Figure 1 shows a screenshot of the website. As observed four webpages are accessible from the frontpage:

1) **Home page.** Includes a general description of the project along with the date of the latest update of the web and a link to a working paper with the main highlights and strategic maps for publishers by fields.

2) **Indicators.** Here the user may choose to consult the tables of indicators for publishers by fields and disciplines or to visit the profile of a given publisher.

*Figure 1. Screenshot of the BiP project website*

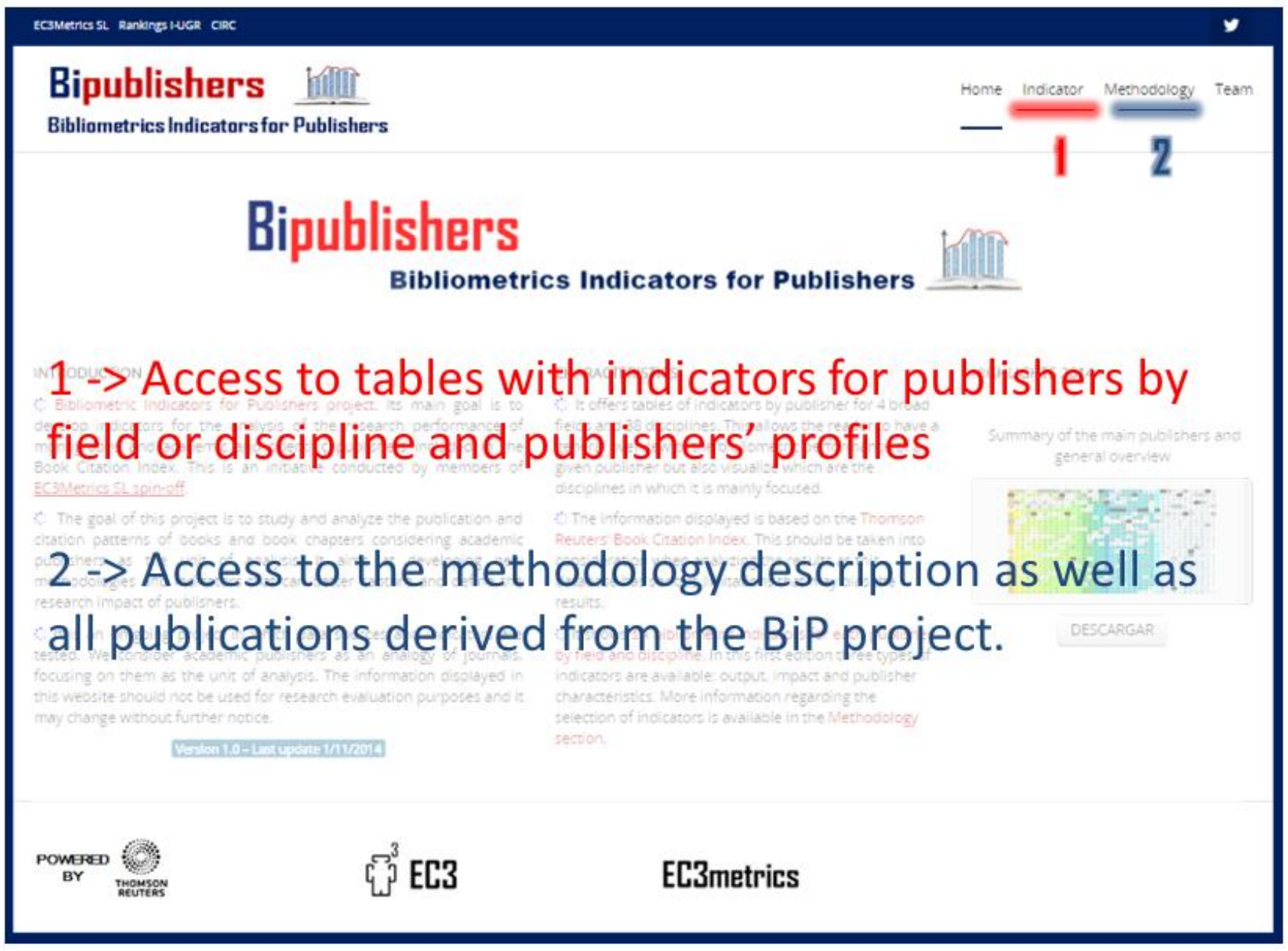

http://bipublishers.es

3) **Methodology.** A submenu unfolds with two options: data and indicators or publications. Data and indicators describes the methodology followed for the data retrieval and normalization process and a definition of the indicators employed. It also gives access to an xls file with the construction of fields and disciplines based on an aggregation of the subject categories employed by the BKCI. The Publications section displays all the research output of the BiP project team with regard to the use of monographs for research evaluation purposes.

4) **Team.** Here we include all team members along with a brief bio for each of them and links to their personal website, Twitter account and Google Scholar Citations profile.

We consider this project to be of the interest of the following parties:

- **Bibliometricians.** Developing bibliometric indicators based on academic publishers remains a major challenge to this research community, who have not been able to properly adapt their research evaluation toolbox to the communication and recognition patterns exerted by means of these publication types. This is an important flaw in fields such as the Social Sciences and the Arts & Humanities, where they are key communication channels for researchers.

- **Librarians.** The results derived from this project may help them to decide on the adequacy of the indicators as well as the different bibliometric database as an answer to their users' demands. Indeed, they often ignore the exact content and scope of the databases they are offered to subscribe to, as well as the audience and regard publishers have and how these may be pertinent to their patrons' demands.

- **Scientific publishers.** This project analyzes publishers as its main unit of analysis. The information thrown by the indicators displayed will allow publishers to have a fair image of the relevance or role played by them as well as by their competitors in the different fields and disciplines displayed.

- **Prospective authors.** Learning which the main publishers in each discipline and field are is of extreme usefulness in order to help them to decide which publisher can better capture the interest of their target audience.

- **Evaluation agencies.** Books and book chapters are currently disregarded or considered as secondary by most evaluation agencies. The development of indicators that can rigorously capture the impact of these publication types will serve to acknowledge the work of scholars who choose these venues as their main communication channel.

## 3. Material and methods

In this section we provide detailed account of the BKCI. We indicate how fields and disciplines were constructed along with the data normalization process followed with the publishers. Finally, we  define the indicators shown.

### 3.1 Description of the Book Citation Index and construction of fields and disciplines

All results shown are based on the web version of the BKCI back in April 2014. The time period covered is 2009-2013. For this period 482,470 records where retrieved, distributed in 14 different document types. As it occurs with the rest of Thomson Reuters' citation indexes, a record may have more than one document type. Among others it includes the types 'book' and 'book chapter'. Only records tagged as 'book' or 'book chapter' were included in the analysis.

*Figure 2. Distribution of the Book Citation Index by document type. Time period 2009-2013.*

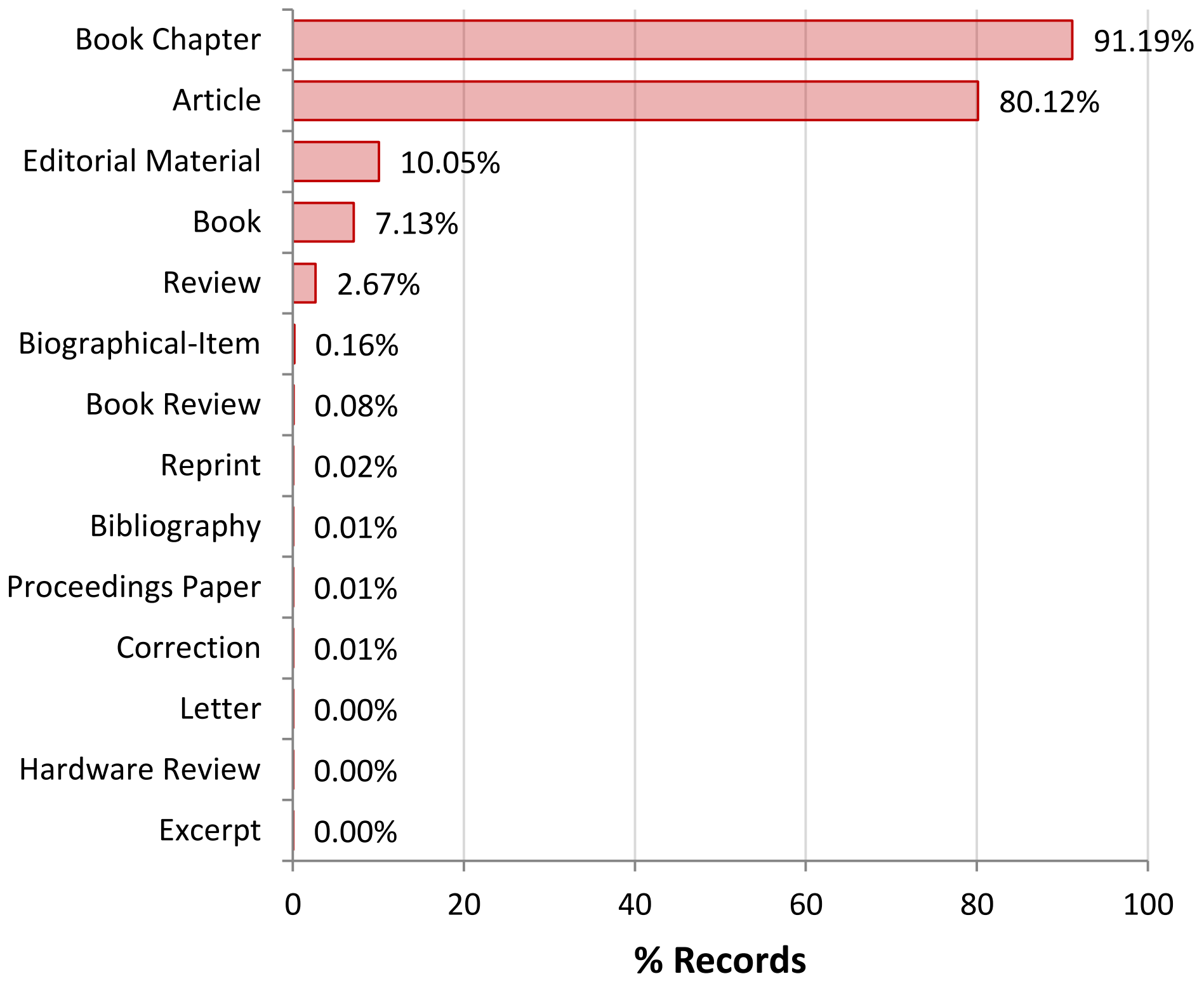

Regarding the construction of fields, this was made through the aggregation of Web of Science subject categories as presented in the BKCI. Unlike to what occurs with journals, books are individually assigned to one or more categories, meaning that a single publisher may have (and usually has) their output distributed among different categories. The aggregation of subject categories for fields and disciplines is available in Appendix A.

## 3.2 Data processing and normalization of publishers

For each record we processed the bibliographic fields as shown in table 1. The field PU was processed separately and normalized manually. We identified 342 different publishers although 254 were finally processed.

*Table 1. Bibliographic fields processed for each record from the Book Citation Index.*

| Acronym | Name | Information |
|---|---|---|
| AU | Author information | Author and affiliation data |
| UT | Accession number | Accession number as identified by Thomson Reuters' Book Citation Index (UT) |
| PT | Publication type | Publication type. Two publication types are identified in the Book Citation Index: S (serials) and B (books). Only records indexed as B were included for the calculation of the indicators. |
| BD | Bibliographic data | Title, source and series information for each record. |
| DT | Document type | 14 document types were found in the Book Citation Index (see figure 1). Only records indexed as books or book chapters were included for the calculation of the indicators. |
| AF | Affiliation | Affiliation and reprint address. |
| IN | Indicators | Number of pages and citations in Web of Science Core Collection and all Web of Science. |
| PU | Publisher | Normalized publisher name, publisher variants and addresses. |
| NR | Reference ID | ISSN or ISBN number of each record. |
| PY | Publication year | Publication year of each record. |
| WC | WoS Category | WoS categories aggregated by disciplines and fields as shown in Appendix A. |

Thomson Reuters provides a list of 499 publishers (http://wokinfo.com/mbl/publishers/), however, many errors were detected in this list. For instance, in the case of Elsevier, the following variants were found:

ACADEMIC PRESS LTD-ELSEVIER SCIENCE LTD
ELSEVIER
ELSEVIER ACADEMIC PRESS INC
ELSEVIER BUTTERWORTH-HEINEMANN
ELSEVIER NORTH HOLLAND
ELSEVIER SCIENCE BV
ELSEVIER SCIENCE LTD
ELSEVIER SCIENCE PUBLISHERS BV BIOMEDICAL DIVISION
ELSEVIER SCIENTIFIC PUBL CO
ELSEVIER/NORTH-HOLLAND
JAI-ELSEVIER LTD
JAI-ELSEVIER SCI BV
JAI-ELSEVIER SCIENCE INC
NORTH HOLLAND, ELSEVIER SCIENCE PUBL BV
PERGAMON-ELSEVIER SCIENCE LTD

To avoid this dispersion, an independent normalization process was conducted manually checking also for further information regarding the publisher (i.e., website). In this process we adopted as a criterion that if a publisher had been acquired by another one, then all its output will be assigned to the latter one. Also, we assigned publisher types, differentiating between two types: 1) commercial and academic publishers, and 2) university presses. The user may filter according to publisher type in the results page at the website.

Next we show two examples of such normalization:

**Example 1: Name variants assigned to Elsevier:**

- ACADEMIC PRESS LTD-ELSEVIER SCIENCE LTD
- ELSEVIER
- ELSEVIER ACADEMIC PRESS INC
- ELSEVIER BUTTERWORTH-HEINEMANN
- ELSEVIER NORTH HOLLAND
- ELSEVIER SCIENCE BV
- ELSEVIER SCIENCE LTD
- ELSEVIER SCIENCE PUBLISHERS BV BIOMEDICAL DIVISION
- ELSEVIER SCIENTIFIC PUBL CO
- GULF PROFESSIONAL PUBL
- GULF PUBL CO
- JAI-ELSEVIER LTD
- MORGAN KAUFMANN PUB INC
- NORTH HOLLAND, ELSEVIER SCIENCE PUBL BV
- PERGAMON-ELSEVIER SCIENCE LTD
- PICKERING & CHATTO PUBLISHERS

**Example 2: name variants assigned to Nottingham University Press**

- NOTTINGHAM UNIVERSITY PRESS
- NOTTINGHAM UNIV PRESS

Publishers are subjected to changes over time, probably more frequently than journal names. In the profile of each publisher all name variants are shown in order to offer a more transparent tool to the user.

Finally a threshold of minimum 5 books or 50 book chapters has been included in order to maintain results stable. Only publishers which surpass such threshold are included in the final tables.

## 3.3 Definition of indicators

Six indicators are provided. In table 2 we include a definition for each of them. The criteria followed for selecting these indicators are to show different aspects of the bibliometric performance of each publishers: output, impact and publisher profile (AI and ED).

*Table 2. Definition of indicators included*

| | Indicator | Acronym | Definition |
|---|---|---|---|
| OUTPUT | Total number of books | PBK | Total number of books published by a given publisher in a certain field or discipline in the last five years. Minimum threshold: 5. |
| | Total number of book chapters | PCH | Total number of book chapters published by a given publisher in a certain field or discipline in the last five years. Minimum threshold: 50. |
| IMPACT | Total number of citations | CIT | Total number of citations received by a given publisher in a certain field or discipline at the time of the data retrieval process. |
| | Field normalized citation score | FNCS | Normalized citations received according to the 'Crown' indicator as defined by Moed et al. (1995). It is interpreted as follows. A publisher with a FNCS of 1 has the same impact as the average of the whole population. Values above one mean that it scores above the average, while values under one underperform in comparison with the global average. |
| PUBLISHER PROFILE | Activity index | AI | Distribution of books in a given field or discipline according to the overall output of a given publisher and in reference to the distribution of the whole Book Citation Index. If the value equals one then, the share of books as of the publisher is the same as the world average. Higher than one means more specialization in the given field. |
| | Percentage of edited items | ED | Share of book chapters which belong to edited books from the total number of book chapters published by a given publisher in a certain field or discipline in the last five years. |

# 4. Results

## 4.1. General overview by fields

From the initial database, 375,655 books and book chapters were finally identified. In table 3 we show the number of records for both document types by fields and for the whole data set. Social Sciences is the field with a higher number of records in overall (33.5%) and by document type (35.7% books and 33.3% book chapters). Next is Science in overall (31.2%) and book chapters (31.7%) and Humanities & Arts in books (29.3%). Engineering & Technology is underrepresented for both document types never reaching 10% of the whole share.

Citations follow a different trend. Here, Science is the field with a higher number of citations for all records (41.0%) and book chapters (56.1%), and it shows similar figures to Social Sciences in regard with books (32% for Science vs. 33.6% for Social Sciences). Engineering & Technology show the lowest values for books (12.7%) and for both document types (13.6%), while Humanities & Arts have only 9.2% of all citations received by book chapters from the whole share.

*Table 3. General output and citation indicators by document type and field according. Time period 2009-2013*

| | BOOKS | | | | BOOK CHAPTERS | | | | ALL | | | |
|---|---|---|---|---|---|---|---|---|---|---|---|---|
| | Total Output | Total Citations | Avg Cit. | Std Dev Cit. | Total Output | Total Citations | Avg Cit. | Std Dev Cit. | Total Output | Total Citations | Avg Cit. | Std Dev Cit. |
| **Science** | 7757 | 42204 | 5.44 | 23.23 | 109559 | 44120 | 0.40 | 2.98 | 117316 | 86324 | 0.74 | 6.75 |
| **Engineering & Tech.** | 2820 | 16729 | 5.93 | 18.08 | 33888 | 11862 | 0.35 | 2.72 | 36708 | 28591 | 0.78 | 5.84 |
| **Social Science** | 10782 | 44231 | 4.10 | 16.04 | 114957 | 15378 | 0.13 | 0.86 | 125739 | 59609 | 0.47 | 4.89 |
| **Humanities** | 8864 | 28672 | 3.23 | 6.79 | 87028 | 7246 | 0.08 | 0.76 | 95892 | 35918 | 0.37 | 2.37 |
| **All Fields** | **30223** | **131836** | **4.36** | **16.59** | **345432** | **78606** | **0.23** | **1.99** | **375655** | **210442** | **0.56** | **5.20** |

## 4.2. Analysis of the indicators by fields

Six indicators were calculated for each publisher by field as described in section 2.3. To understand the relation between them and better interpret their meaning and differences by field, we analyzed their correlation according to the Pearson Coefficient. Table 4 shows the averages and standard deviation for each of the six indicators by field. These indicators use publishers as the main unit of analysis.

*Table 4. Average and standard deviation by fields of the number of books, book chapters, citations, field normalized citation score, activity index and percentage of edited items for publishers in the Book Citation Index. Time period 2009-2013*

| | | PBK | PCH | CIT | FNCS | AI | ED |
|---|---|---|---|---|---|---|---|
| **SCI** | **Average** | 90.36 | 1284.58 | 1016.42 | 0.93 | 1.74 | 0.49 |
| | **Std. Deviation** | 291.86 | 4673.62 | 4160.83 | 1.13 | 1.16 | 0.24 |
| **E&T** | **Average** | 74.14 | 892.46 | 758.03 | 0.76 | 3.86 | 0.50 |
| | **Std. Deviation** | 184.69 | 2162.40 | 2501.15 | 0.50 | 3.29 | 0.22 |
| **SOC** | **Average** | 134.86 | 1439.58 | 743.84 | 0.75 | 1.21 | 0.43 |
| | **Std. Deviation** | 364.58 | 3850.95 | 1866.23 | 0.65 | 0.65 | 0.21 |
| **HUM** | **Average** | 116.53 | 1143.56 | 475.21 | 0.88 | 1.51 | 0.40 |
| | **Std. Deviation** | 281.41 | 2632.49 | 1048.60 | 0.75 | 0.86 | 0.22 |

Table 5 displays Pearson Coefficients between the six indicators for each of the four analyzed fields. As observed, a similar trend is found in all fields. While there is a strong correlation between number of books, number of book chapters and citations, there is almost no correlation with the rest of the indicators. This means that they are showing different aspects of the performance of publishers. Furthermore,

*Table 5. Correlation tables by fields between the number of books, book chapters, citations, field normalized citation score, activity index and percentage of edited items according to the Book Citation Index. Time period 2009-2013*

| SCIENCE | PBK | PCH | CIT | FNCS | AI | ED |
|---|---|---|---|---|---|---|
| PBK | **1.00** | **0.99** | **0.95** | 0.03 | 0.03 | 0.15 |
| PCH | **0.99** | **1.00** | **0.97** | 0.02 | 0.04 | 0.15 |
| CIT | **0.95** | **0.97** | **1.00** | 0.14 | 0.04 | 0.11 |
| FNCS | 0.03 | 0.02 | 0.14 | **1.00** | 0.16 | -0.05 |
| AI | 0.03 | 0.04 | 0.04 | 0.16 | **1.00** | **0.25** |
| ED | 0.15 | 0.15 | 0.11 | -0.05 | **0.25** | **1.00** |

| ENGINEERING & TECHNOLOGY | PBK | PCH | CIT | FNCS | AI | ED |
|---|---|---|---|---|---|---|
| PBK | **1.00** | **1.00** | **0.97** | 0.22 | -0.03 | 0.02 |
| PCH | **1.00** | **1.00** | **0.97** | 0.22 | -0.02 | 0.01 |
| CIT | **0.97** | **0.97** | **1.00** | 0.28 | -0.07 | 0.00 |
| FNCS | 0.22 | 0.22 | 0.28 | **1.00** | **-0.45** | -0.08 |
| AI | -0.03 | -0.02 | -0.07 | **-0.45** | **1.00** | -0.27 |
| ED | 0.02 | 0.01 | 0.00 | -0.08 | -0.27 | **1.00** |

| SOCIAL SCIENCES | PBK | PCH | CIT | FNCS | AI | ED |
|---|---|---|---|---|---|---|
| PBK | **1.00** | **0.99** | **0.88** | 0.15 | 0.09 | 0.04 |
| PCH | **0.99** | **1.00** | **0.88** | 0.14 | 0.10 | 0.06 |
| CIT | **0.88** | **0.88** | **1.00** | **0.45** | 0.06 | 0.00 |
| FNCS | 0.15 | 0.14 | **0.45** | **1.00** | -0.10 | -0.05 |
| AI | 0.09 | 0.10 | 0.06 | -0.10 | **1.00** | 0.10 |
| ED | 0.04 | 0.06 | 0.00 | -0.05 | 0.10 | **1.00** |

| HUMANITIES & ARTS | PBK | PCH | CIT | FNCS | AI | ED |
|---|---|---|---|---|---|---|
| PBK | **1.00** | **0.99** | **0.91** | 0.08 | 0.02 | -0.01 |
| PCH | **0.99** | **1.00** | **0.90** | 0.07 | -0.01 | 0.00 |
| CIT | **0.91** | **0.90** | **1.00** | **0.28** | -0.01 | -0.03 |
| FNCS | 0.08 | 0.07 | **0.28** | **1.00** | -0.14 | **0.34** |
| AI | 0.02 | -0.01 | -0.01 | -0.14 | **1.00** | 0.01 |
| ED | -0.01 | 0.00 | -0.03 | **0.34** | 0.01 | **1.00** |

**Note:** In bold values ≠ 0 with a significance level alpha =0.05

Furthermore, we see that there are important differences between considering the raw number of citations and applying a normalized citation score, emphasizing the importance of considering non-size dependent indicators to measure the research impact of publications.

## 4.3. Main publishers by fields

From the 342 publishers identified, only 126 publishers surpassed the minimum publication threshold described in section 3.2 in at least one of the four fields displayed and 113 made it to at least one of the 38 disciplines. Figure 3 shows the number of publishers by field and table 6 does the same for disciplines.

*Figure 3. Distribution of publishers by field and publisher type in the Book Citation Index. Time period 2009-2013*

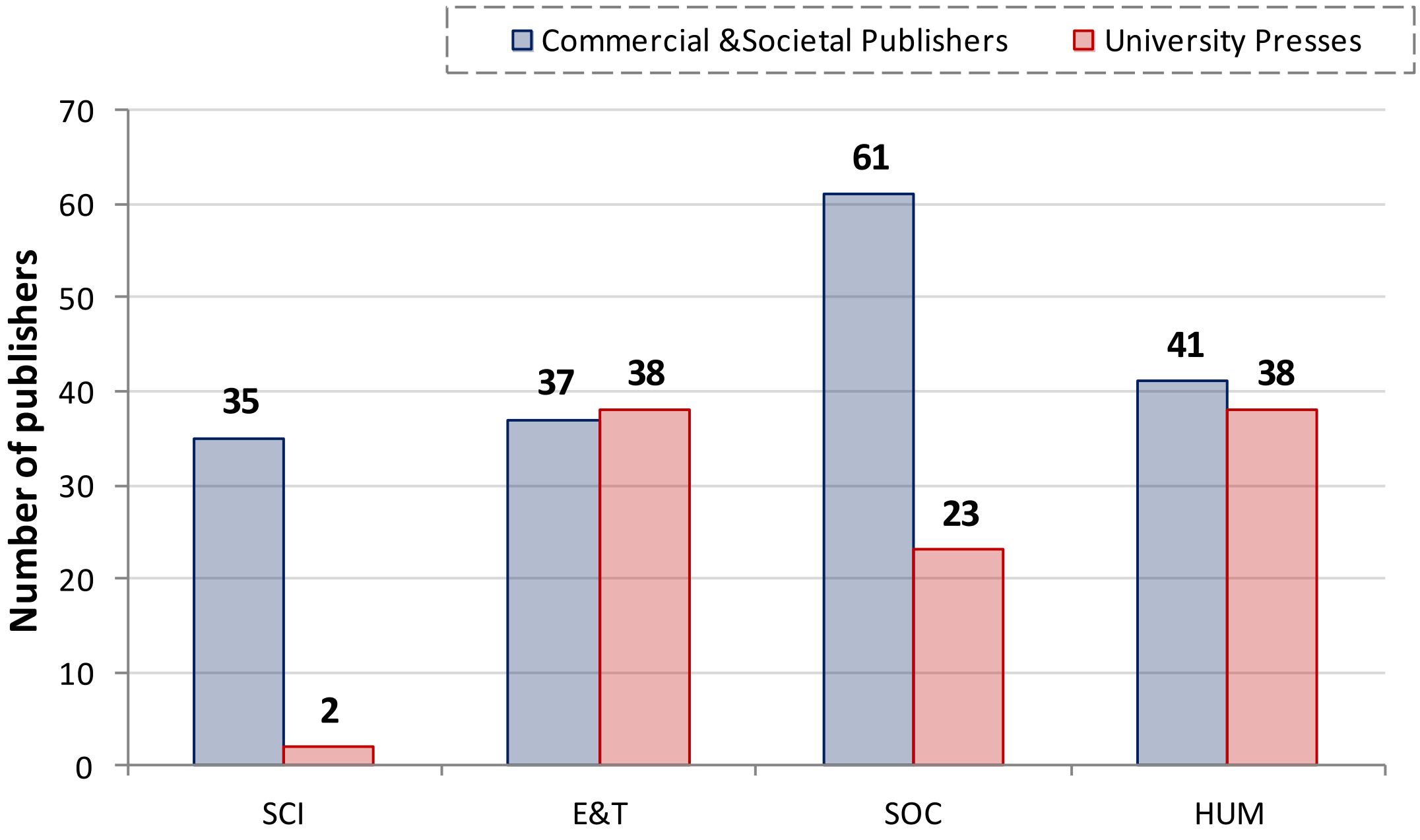

*Table 6. Distribution of publishers by discipline and publisher – Commercial & Society Publishers or University Press. type in the Book Citation Index. Time period 2009-2013*

| Discipline | Nr of Commercial & Society Publishers | | Nr of University Press Publishers | | Discipline | Nr of Commercial & Society Publishers | | Nr of University Press Publishers | |
|---|---|---|---|---|---|---|---|---|---|
| Agricultural Sciences | 9 | 82% | 2 | 18% | History & Philosophy of Science | 18 | 56% | 14 | 44% |
| Anthropology | 4 | 25% | 12 | 75% | Humanities, Miscellaneus | 12 | 43% | 16 | 57% |
| Applied Physics and Chemistry | 9 | 82% | 2 | 18% | Information Science | 8 | 80% | 2 | 20% |
| Archeology | 2 | 40% | 3 | 60% | Language & Linguistics | 11 | 61% | 7 | 39% |
| Architecture & Urban Studies | 19 | 56% | 15 | 44% | Law, Criminology & Penology | 26 | 48% | 28 | 52% |
| Area & Cultural Studies | 26 | 39% | 41 | 61% | Literature | 9 | 28% | 23 | 72% |
| Arts | 7 | 33% | 14 | 67% | Materials Science | 10 | 91% | 1 | 8% |
| Biological Sci - Animals & Plants | 19 | 83% | 4 | 17% | Mathemathics | 15 | 83% | 3 | 17% |
| Biological Sci - Humans | 18 | 82% | 4 | 18% | Medicine & Clinical Medicine | 29 | 73% | 11 | 27% |
| Chemistry | 11 | 92% | 1 | 8% | Mol Biology and Biochem | 13 | 93% | 1 | 7% |
| Communication | 11 | 44% | 14 | 56% | Pharmacology & Pharmacy | 6 | 86% | 1 | 14% |
| Computer Science | 10 | 83% | 2 | 17% | Philosophy & Ethics | 9 | 41% | 13 | 59% |
| Earth & Enviromental Sciences | 29 | 67% | 14 | 33% | Physics & Astronomy | 10 | 83% | 2 | 17% |
| Economics, Bussiness & Mgmnt | 26 | 65% | 14 | 35% | Political Science | 22 | 47% | 25 | 53% |
| Education | 17 | 61% | 11 | 39% | Psychology | 14 | 78% | 4 | 22% |
| Engineering | 18 | 90% | 2 | 10% | Social Sciences, Interdisciplinary | 15 | 43% | 20 | 57% |
| Folklore & Ethnic Studies | 5 | 26% | 14 | 74% | Social Work | 5 | 83% | 1 | 17% |
| Geography | 19 | 48% | 21 | 52% | Sociology | 10 | 48% | 11 | 52% |
| History | 11 | 30% | 26 | 70% | Technology | 22 | 92% | 2 | 8% |

When analyzing the whole output of publishers (books and book chapters), we observe that 20 publishers represent roughly 2/3 of the overall output for the analyzed time period (table 7). In fact, between the three most productive (Springer, Palgrave MacMillan and Routledge) they account for 35.4% of the whole share. Regarding the number of citations received, book chapters show larger figures in general, however, the average number of citations is much higher for books. In any case, the standard deviations are very high, signifying a skewed distribution. Appendix B shows these indicators for all publishers.

*Table 7. 20 most productive publishers according to both document types (books and book chapters) in the Book Citation Index. Time period 2009-2013*

| PUBLISHER | BOOKS | | | | BOOK CHAPTERS | | | | ALL | | | |
|---|---|---|---|---|---|---|---|---|---|---|---|---|
| | Total | Cit | Avg Cit | Std Dev Cit | Total | Cit | Avg Cit | Std Dev Cit | Total | Cit | Avg Cit | Std Dev Cit |
| Springer | 3799 | 16013 | 4.22 | 12.61 | 56193 | 33398 | 0.59 | 3.45 | 59992 | 49411 | 0.82 | 4.69 |
| Palgrave Macmillan | 4213 | 10640 | 2.53 | 4.47 | 41093 | 2712 | 0.07 | 0.41 | 45306 | 13352 | 0.29 | 1.59 |
| Routledge | 2176 | 8991 | 4.13 | 12.62 | 25335 | 2594 | 0.1 | 0.49 | 27511 | 11585 | 0.42 | 3.74 |
| Cambridge Univ Press | 1755 | 11973 | 6.82 | 21.22 | 15988 | 1624 | 0.1 | 0.89 | 17743 | 13597 | 0.77 | 7.02 |
| Elsevier | 883 | 7429 | 8.41 | 32.55 | 15739 | 2642 | 0.17 | 0.88 | 16622 | 10071 | 0.61 | 7.77 |
| Nova Science Publishers | 1336 | 828 | 0.62 | 2.11 | 14391 | 3125 | 0.22 | 0.89 | 15727 | 3953 | 0.25 | 1.05 |
| Edward Elgar | 997 | 2384 | 2.39 | 5.06 | 12301 | 1635 | 0.13 | 0.56 | 13298 | 4019 | 0.3 | 1.6 |
| Information Age Publishing | 443 | 328 | 0.74 | 1.96 | 5819 | 1302 | 0.22 | 2.07 | 6262 | 1630 | 0.26 | 2.06 |
| Princeton Univ Press | 599 | 11073 | 18.49 | 51.03 | 5608 | 181 | 0.03 | 0.26 | 6207 | 11254 | 1.81 | 16.75 |
| Univ of California Press | 552 | 3797 | 6.88 | 13.9 | 5601 | 292 | 0.05 | 0.42 | 6153 | 4089 | 0.66 | 4.61 |
| MIT Press | 396 | 3448 | 8.71 | 19.8 | 4233 | 377 | 0.09 | 0.57 | 4629 | 3825 | 0.83 | 6.29 |
| De Gruyter | 283 | 832 | 2.94 | 8.47 | 3626 | 1099 | 0.3 | 2.71 | 3909 | 1931 | 0.49 | 3.53 |
| Taylor & Francis | 294 | 902 | 3.07 | 6.42 | 3525 | 291 | 0.08 | 0.48 | 3819 | 1193 | 0.31 | 2 |
| Univ of Pennsylvania Press | 391 | 2277 | 5.82 | 6.99 | 3306 | 478 | 0.14 | 0.87 | 3697 | 2755 | 0.75 | 2.98 |
| CRC Press | 239 | 1628 | 6.81 | 17.11 | 3422 | 2993 | 0.87 | 8.69 | 3661 | 4621 | 1.26 | 9.58 |
| Wiley-Blackwell | 226 | 424 | 1.88 | 8.67 | 3181 | 425 | 0.13 | 0.94 | 3407 | 849 | 0.25 | 2.44 |
| McGill-Queens Univ Press | 284 | 514 | 1.81 | 2.64 | 3055 | 100 | 0.03 | 0.25 | 3339 | 614 | 0.18 | 0.95 |
| Woodhead Publishing | 214 | 583 | 2.72 | 6.64 | 2784 | 435 | 0.16 | 0.56 | 2998 | 1018 | 0.34 | 1.96 |
| Univ of North Carolina Press | 269 | 1906 | 7.09 | 6.79 | 2377 | 75 | 0.03 | 0.41 | 2646 | 1981 | 0.75 | 3.06 |
| Brill | 173 | 234 | 1.35 | 3.21 | 2330 | 118 | 0.05 | 0.28 | 2503 | 352 | 0.14 | 0.94 |

Table 8 shows the top ten publishers with a higher number of monographs for each of the four fields. We observe two publishers (Springer and Cambridge University Press) are present in the top ten for all fields, three publishers are present in three fields, always with the exception of Humanities & Arts or Engineering & Technology (Nova Science Publishers, Routledge and Palgrave MacMillan). Then we observe two groups formed by four and three publishers which are either present in Science and Engineering & Technology (Elsevier, Wiley-Blackwell, CRC Press and MIT Press), or in Social Sciences and Humanities & Arts (Edward Elgar, Princeton University Press and University of California Press). Finally, 9 publishers are only present in the top ten of one of the four fields.

Table 8. Top ten publishers with a higher number of books for Sciences, Social Sciences, Engineering & Technology, and Humanities & Arts according to the Book Citation Index. Time period 2009-2013

| SCIENCE | PBK |
|---|---|
| **Springer** | 2446 |
| **Nova Science Publishers** | 961 |
| **Elsevier** | 538 |
| **Cambridge University Press** | 417 |
| **Routledge** | 361 |
| **Palgrave Macmillan** | 211 |
| **Wiley-Blackwell** | 148 |
| **CRC Press** | 145 |
| **MIT Press** | 145 |
| **CABI** | 140 |

| ENGINEERING & TECHNOLOGY | PBK |
|---|---|
| **Springer** | 1054 |
| **Elsevier** | 387 |
| **Nova Science Publishers** | 267 |
| **Woodhead Publishing** | 192 |
| **Artech House** | 142 |
| **CRC Press** | 126 |
| **Cambridge University Press** | 99 |
| **Pan-Stanford** | 79 |
| **Wiley-Blackwell** | 49 |
| **MIT Press** | 47 |

| SOCIAL SCIENCES | PBK |
|---|---|
| **Palgrave Macmillan** | 2680 |
| **Routledge** | 1540 |
| **Edward Elgar** | 814 |
| **Springer** | 787 |
| **Cambridge University Press** | 513 |
| **Information Age Publishing** | 415 |
| **Princeton University Press** | 314 |
| **Nova Science Publishers** | 249 |
| **World Bank** | 239 |
| **University of California Press** | 223 |

| HUMANITIES & ARTS | PBK |
|---|---|
| **Palgrave Macmillan** | 2108 |
| **Cambridge University Press** | 1004 |
| **Routledge** | 748 |
| **Springer** | 383 |
| **Princeton University Press** | 339 |
| **University of Pennsylvania Press** | 292 |
| **University of California Press** | 290 |
| **Edward Elgar** | 266 |
| **University of North Carolina Press** | 226 |
| **Edinburgh University Press** | 184 |

## 5. Concluding remarks

The BiP project attempts to address a well-known limitation of bibliometric studies, which is the lack of reliable indicators for analyzing the research performance of monographs and book chapters. An issue which affects especially research evaluation in the fields of Social Sciences and Humanities & Arts. It focuses on the analysis of publishers following a similar analogy to that adopted by Garfield (1972) when analyzing journals instead of articles themselves. The data is retrieved from Thomson Reuters' Book Citation Index, which has been reported to have strong biases (Torres-Salinas et al., 2013; 2014a), something which should be kept in mind when analyzing the results offered.

Such results are displayed througout six bibliometric indicators, three of them (number of books, number of chapters and number of citations) correlate strongly with each other while the other three show no correlation with the rest (field normalized citation score, activity index and percentage of edited items). The list of publishers with a higher presence in the BKCI is consistent with the results shown by Torres-Salinas and colleagues (2014a). Although there is an increase in the number of publishers, still there is a high concentration of the main scientific publishers which represent most of the BKCI.

All results of the BiP project are available at http://bipublishers.es. This project is a research initiative of EC3Metrics SL.

## Appendix A. Construction of fields and disciplines according to the Web os Science subject categories

| ENGINEERING & TECHNOLOGY | |
|---|---|
| DISCIPLINE | WOS CATEGORY |
| **Computer Science** | Computer Science, Artificial Intelligence |
| | Computer Science, Cybernetics |
| | Computer Science, Hardware & Architecture |
| | Computer Science, Information Systems |
| | Computer Science, Interdisciplinary Applications |
| | Computer Science, Software Engineering |
| | Computer Science, Theory & Methods |
| **Engineering** | Engineering, Aerospace |
| | Engineering, Biomedical |
| | Engineering, Chemical |
| | Engineering, Civil |
| | Engineering, Electrical & Electronic |
| | Engineering, Environmental |
| | Engineering, Geological |
| | Engineering, Industrial |
| | Engineering, Manufacturing |
| | Engineering, Marine |
| | Engineering, Mechanical |
| | Engineering, Multidisciplinary |
| | Engineering, Ocean |
| | Engineering, Petroleum |
| | Metallurgy & Metallurgical Engineering |
| **Materials Science** | Materials Science, Biomaterials |
| | Materials Science, Ceramics |
| | Materials Science, Characterization & Testing |
| | Materials Science, Coatings & Films |
| | Materials Science, Composites |
| | Materials Science, Multidisciplinary |
| | Materials Science, Paper & Wood |
| | Materials Science, Textiles |
| **Technology** | Automation & Control Systems |
| | Construction & Building Technology |
| | Ergonomics |
| | Imaging Science & Photographic Technology |
| | Medical Informatics |

| ENGINEERING & TECHNOLOGY | |
|---|---|
| | Medical Laboratory Technology |
| | Nanoscience & Nanotechnology |
| | Nuclear Science & Technology |
| | Remote Sensing |
| | Robotics |
| | Telecommunications |
| | Transportation |
| | Transportation Science & Technology |

| HUMANITIES & ARTS<br>DISCIPLINE | <br>WOS CATEGORY |
|---|---|
| **Archeology** | Archaeology |
| **Architecture & Urban Studies** | Architecture |
| **Area & Cultural Studies** | Asian Studies |
| | Cultural Studies |
| **Arts** | Art |
| | Dance |
| | Music |
| | Theater |
| **Folklore & Ethnic Studies** | Ethnic Studies |
| | Folklore |
| **Geography** | Demography |
| **History** | History |
| **History & Philosophy of Science** | History & Philosophy Of Science |
| **Humanities, Miscellaneus** | Classics |
| | Humanities, Multidisciplinary |
| | Medieval & Renaissance Studies |
| | Religion |
| **Languague & Linguistics** | Language & Linguistics |
| | Linguistics |
| **Law, Criminology & Penology** | Law |
| **Literature** | Literary Reviews |
| | Literary Theory & Criticism |
| | Literature |
| | Literature, African, Australian, Canadian |
| | Literature, American |
| | Literature, British Isles |
| | Literature, German, Dutch, Scandinavian |
| | Literature, Romance |

| HUMANITIES & ARTS | |
|---|---|
| | Literature, Slavic<br>Poetry |
| **Philosophy & Ethics** | Ethics<br>Logic<br>Philosophy |

| SCIENCE<br>DISCIPLINE | WOS CATEGORY |
|---|---|
| **Agricultural Sciences** | Agricultural Economics & Policy<br>Agricultural Engineering<br>Agriculture, Dairy & Animal Science<br>Agriculture, Multidisciplinary<br>Agronomy<br>Horticulture |
| **Applied Physics and Chemistry** | Acoustics<br>Instruments & Instrumentation<br>Mechanics<br>Optics<br>Physics, Applied |
| **Biological Sciences - Animals & Plants** | Biodiversity Conservation<br>Ecology<br>Entomology<br>Fisheries<br>Food Science & Technology<br>Marine & Freshwater Biology<br>Mycology<br>Ornithology<br>Plant Sciences<br>Veterinary Sciences<br>Zoology |
| **Biological Sciences - Humans** | Anatomy & Morphology<br>Andrology<br>Biology<br>Biophysics<br>Endocrinology & Metabolism<br>Evolutionary Biology<br>Hematology<br>Immunology<br>Mathematical & Computational Biology<br>Microbiology<br>Microscopy<br>Multidisciplinary Sciences<br>Neuroimaging<br>Neurosciences<br>Parasitology<br>Physiology |

| SCIENCE | |
|---|---|
| | Reproductive Biology<br>Toxicology<br>Virology |
| **Chemistry** | Chemistry, Analytical<br>Chemistry, Applied<br>Chemistry, Inorganic & Nuclear<br>Chemistry, Medicinal<br>Chemistry, Multidisciplinary<br>Chemistry, Organic<br>Chemistry, Physical<br>Crystallography<br>Electrochemistry<br>Polymer Science |
| **Earth & Enviromental Sciences** | Energy & Fuels<br>Environmental Sciences<br>Environmental Studies<br>Forestry<br>Geochemistry & Geophysics<br>Geography, Physical<br>Geology<br>Geosciences, Multidisciplinary<br>Limnology<br>Meteorology & Atmospheric Sciences<br>Mineralogy<br>Mining & Mineral Processing<br>Oceanography<br>Paleontology<br>Soil Science<br>Water Resources |
| **Mathemathics** | Mathematics<br>Mathematics, Applied<br>Mathematics, Interdisciplinary Applications<br>Operations Research & Management Science<br>Statistics & Probability |
| **Medicine & Clinical Medicine** | Allergy<br>Anesthesiology<br>Audiology & Speech-Language Pathology<br>Cardiac & Cardiovascular Systems<br>Clinical Neurology<br>Critical Care Medicine<br>Dentistry, Oral Surgery & Medicine<br>Dermatology<br>Emergency Medicine<br>Gastroenterology & Hepatology<br>Genetics & Heredity<br>Geriatrics & Gerontology<br>Gerontology |

| SCIENCE | |
| --- | --- |
| | Health Care Sciences & Services<br>Health Policy & Services<br>Infectious Diseases<br>Integrative & Complementary Medicine<br>Medical Ethics<br>Medicine, General & Internal<br>Medicine, Legal<br>Medicine, Research & Experimental<br>Nursing<br>Nutrition & Dietetics<br>Obstetrics & Gynecology<br>Oncology<br>Ophthalmology<br>Orthopedics<br>Otorhinolaryngology<br>Pathology<br>Pediatrics<br>Peripheral Vascular Disease<br>Primary Health Care<br>Psychiatry<br>Public, Environmental & Occupational Health<br>Radiology, Nuclear Medicine & Medical Imaging<br>Rehabilitation<br>Respiratory System<br>Rheumatology<br>Social Sciences, Biomedical<br>Spectroscopy<br>Sport Sciences<br>Substance Abuse<br>Surgery<br>Transplantation<br>Tropical Medicine<br>Urology & Nephrology |
| **Molecular Biology and Biochemistry** | Biochemical Research Methods<br>Biochemistry & Molecular Biology<br>Biotechnology & Applied Microbiology<br>Cell & Tissue Engineering<br>Cell Biology<br>Developmental Biology |
| **Pharmacology & Pharmacy** | Pharmacology & Pharmacy |
| **Physics & Astronomy** | Astronomy & Astrophysics<br>Physics, Atomic, Molecular & Chemical<br>Physics, Condensed Matter<br>Physics, Fluids & Plasmas<br>Physics, Mathematical<br>Physics, Multidisciplinary<br>Physics, Nuclear |

| SCIENCE | |
|---|---|
| | Physics, Particles & Fields<br>Thermodynamics |
| **Psychology** | Behavioral Sciences<br>Psychology<br>Psychology, Applied<br>Psychology, Biological<br>Psychology, Clinical<br>Psychology, Developmental<br>Psychology, Experimental<br>Psychology, Mathematical<br>Psychology, Multidisciplinary<br>Psychology, Psychoanalysis<br>Psychology, Social |

| SOCIAL SCIENCES | |
|---|---|
| DISCIPLINE | WOS CATEGORY |
| **Anthropology** | Anthropology |
| **Architecture & Urban Studies** | Urban Studies |
| **Area & Cultural Studies** | Area Studies<br>Social Issues |
| **Communication** | Communication<br>Film, Radio, Television |
| **Economics, Bussiness & Managment** | Business<br>Business, Finance<br>Economics<br>Industrial Relations & Labor<br>Management |
| **Education** | Education & Educational Research<br>Education, Scientific Disciplines<br>Education, Special<br>Psychology, Educational |
| **Geography** | Geography<br>Planning & Development |
| **History & Philosophy of Science** | History Of Social Sciences |
| **Information Science & Library Science** | Information Science & Library Science |
| **Law, Criminology & Penology** | Criminology & Penology |
| **Political Science & International Relations** | International Relations<br>Political Science<br>Public Administration |
| **Social Sciences, Interdisciplinary** | Family Studies<br>Hospitality, Leisure, Sport & Tourism<br>Social Sciences, Interdisciplinary<br>Social Sciences, Mathematical Methods<br>Women's Studies |
| **Social Work** | Social Work |
| **Sociology** | Sociology |

## Appendix B. Main bibliometric indicators for publishers identified in the BKCI

| | Book | Cit | Avg Cit | Std Dev Cit | Book Chapter | Cit | Avg Cit | Std Dev Cit | All | Cit | Avg Cit | Std Dev Cit |
|---|---|---|---|---|---|---|---|---|---|---|---|---|
| Springer | 3799 | 16013 | 4.22 | 12.61 | 56193 | 33398 | 0.594 | 3.446 | 59992 | 49411 | 0.82 | 4.69 |
| Palgrave Macmillan | 4213 | 10640 | 2.53 | 4.47 | 41093 | 2712 | 0.066 | 0.413 | 45306 | 13352 | 0.29 | 1.59 |
| Routledge | 2176 | 8991 | 4.13 | 12.62 | 25335 | 2594 | 0.102 | 0.492 | 27511 | 11585 | 0.42 | 3.74 |
| Cambridge University Press | 1755 | 11973 | 6.82 | 21.22 | 15988 | 1624 | 0.102 | 0.887 | 17743 | 13597 | 0.77 | 7.02 |
| Elsevier | 883 | 7429 | 8.41 | 32.55 | 15739 | 2642 | 0.168 | 0.882 | 16622 | 10071 | 0.61 | 7.77 |
| Nova Science Publishers | 1336 | 828 | 0.62 | 2.11 | 14391 | 3125 | 0.217 | 0.886 | 15727 | 3953 | 0.25 | 1.05 |
| Edward Elgar | 997 | 2384 | 2.39 | 5.06 | 12301 | 1635 | 0.133 | 0.556 | 13298 | 4019 | 0.30 | 1.60 |
| Information Age Publishing | 443 | 328 | 0.74 | 1.96 | 5819 | 1302 | 0.224 | 2.066 | 6262 | 1630 | 0.26 | 2.06 |
| Princeton University Press | 599 | 11073 | 18.49 | 51.03 | 5608 | 181 | 0.032 | 0.259 | 6207 | 11254 | 1.81 | 16.75 |
| University of California Press | 552 | 3797 | 6.88 | 13.90 | 5601 | 292 | 0.052 | 0.418 | 6153 | 4089 | 0.66 | 4.61 |
| MIT Press | 396 | 3448 | 8.71 | 19.80 | 4233 | 377 | 0.089 | 0.569 | 4629 | 3825 | 0.83 | 6.29 |
| De Gruyter | 283 | 832 | 2.94 | 8.47 | 3626 | 1099 | 0.303 | 2.714 | 3909 | 1931 | 0.49 | 3.53 |
| Taylor & Francis | 294 | 902 | 3.07 | 6.42 | 3525 | 291 | 0.083 | 0.476 | 3819 | 1193 | 0.31 | 2.00 |
| University of Pennsylvania Press | 391 | 2277 | 5.82 | 6.99 | 3306 | 478 | 0.145 | 0.865 | 3697 | 2755 | 0.75 | 2.98 |
| CRC Press | 239 | 1628 | 6.81 | 17.11 | 3422 | 2993 | 0.875 | 8.690 | 3661 | 4621 | 1.26 | 9.58 |
| Wiley-Blackwell | 226 | 424 | 1.88 | 8.67 | 3181 | 425 | 0.134 | 0.937 | 3407 | 849 | 0.25 | 2.44 |
| McGill-Queens University Press | 284 | 514 | 1.81 | 2.64 | 3055 | 100 | 0.033 | 0.246 | 3339 | 614 | 0.18 | 0.95 |
| Woodhead Publishing | 214 | 583 | 2.72 | 6.64 | 2784 | 435 | 0.156 | 0.561 | 2998 | 1018 | 0.34 | 1.96 |
| University of North Carolina Press | 269 | 1906 | 7.09 | 6.79 | 2377 | 75 | 0.032 | 0.409 | 2646 | 1981 | 0.75 | 3.06 |
| Brill | 173 | 234 | 1.35 | 3.21 | 2330 | 118 | 0.051 | 0.283 | 2503 | 352 | 0.14 | 0.94 |
| University of Illinois Press | 235 | 733 | 3.12 | 4.63 | 2245 | 61 | 0.027 | 0.206 | 2480 | 794 | 0.32 | 1.70 |
| World Bank | 261 | 2235 | 8.56 | 37.05 | 2077 | 164 | 0.079 | 0.411 | 2338 | 2399 | 1.03 | 12.65 |
| Edinburgh University Press | 241 | 340 | 1.41 | 3.06 | 2091 | 33 | 0.016 | 0.155 | 2332 | 373 | 0.16 | 1.08 |
| CABI | 169 | 536 | 3.17 | 5.52 | 2066 | 612 | 0.296 | 1.023 | 2235 | 1148 | 0.51 | 1.96 |
| Artech House | 157 | 514 | 3.27 | 6.46 | 1952 | 341 | 0.175 | 1.682 | 2109 | 855 | 0.41 | 2.52 |
| University Press of Mississippi | 168 | 290 | 1.73 | 2.66 | 1915 | 56 | 0.029 | 0.257 | 2083 | 346 | 0.17 | 0.92 |
| Pennsylvania State University | 186 | 680 | 3.66 | 3.84 | 1807 | 140 | 0.077 | 0.474 | 1993 | 820 | 0.41 | 1.63 |
| University Press of Kentucky | 146 | 291 | 1.99 | 3.18 | 1704 | 52 | 0.031 | 0.238 | 1850 | 343 | 0.19 | 1.06 |
| SLACK Incorporated | 77 | 146 | 1.90 | 9.84 | 1677 | 130 | 0.078 | 0.414 | 1754 | 276 | 0.16 | 2.12 |
| Demos Medical Publishing | 65 | 46 | 0.71 | 1.64 | 1667 | 100 | 0.060 | 0.294 | 1732 | 146 | 0.08 | 0.44 |
| Chandos | 165 | 386 | 2.34 | 3.63 | 1551 | 43 | 0.028 | 0.233 | 1716 | 429 | 0.25 | 1.33 |
| Fordham University Press | 161 | 275 | 1.71 | 3.25 | 1423 | 85 | 0.060 | 0.529 | 1584 | 360 | 0.23 | 1.25 |
| Intellect Ltd. | 118 | 227 | 1.92 | 3.43 | 1372 | 31 | 0.023 | 0.163 | 1490 | 258 | 0.17 | 1.10 |
| Brookings Institute | 136 | 681 | 5.01 | 7.98 | 1290 | 327 | 0.253 | 2.045 | 1426 | 1008 | 0.71 | 3.43 |
| Pan-Stanford | 111 | 407 | 3.67 | 14.20 | 1255 | 135 | 0.108 | 0.473 | 1366 | 542 | 0.40 | 4.17 |
| University of Virginia Press | 135 | 466 | 3.45 | 3.97 | 1145 | 36 | 0.031 | 0.202 | 1280 | 502 | 0.39 | 1.67 |
| University of Nebraska Press | 100 | 14 | 0.14 | 0.47 | 1132 | 64 | 0.057 | 0.408 | 1232 | 78 | 0.06 | 0.41 |
| Multilingual Matters | 119 | 594 | 4.99 | 9.68 | 1098 | 301 | 0.274 | 0.990 | 1217 | 895 | 0.74 | 3.45 |
| Australia National University Press | 102 | 254 | 2.49 | 3.46 | 1071 | 193 | 0.180 | 0.577 | 1173 | 447 | 0.38 | 1.33 |
| IGI Global | 75 | 79 | 1.05 | 7.04 | 1081 | 40 | 0.037 | 0.244 | 1156 | 119 | 0.10 | 1.82 |
| Louisiana State University Press | 143 | 388 | 2.71 | 3.21 | 949 | 9 | 0.009 | 0.097 | 1092 | 397 | 0.36 | 1.48 |
| CSIRO Publishers | 75 | 410 | 5.47 | 9.86 | 954 | 112 | 0.117 | 0.614 | 1029 | 522 | 0.51 | 3.05 |
| University of Minnesota Press | 95 | 784 | 8.25 | 13.31 | 919 | 63 | 0.069 | 0.402 | 1014 | 847 | 0.84 | 4.72 |
| University of British Columbia Press | 94 | 364 | 3.87 | 4.24 | 887 | 111 | 0.125 | 0.745 | 981 | 475 | 0.48 | 1.85 |
| University Press of Colorado | 67 | 231 | 3.45 | 3.73 | 877 | 75 | 0.086 | 0.502 | 944 | 306 | 0.32 | 1.40 |
| World Scientific Publishing Co Pte Ltd | 71 | 97 | 1.37 | 5.97 | 841 | 48 | 0.057 | 0.487 | 912 | 145 | 0.16 | 1.75 |
| Caister Academic Press | 64 | 104 | 1.63 | 2.76 | 840 | 1037 | 1.235 | 2.277 | 904 | 1141 | 1.26 | 2.31 |
| Royal Society of Chemistry | 63 | 951 | 15.10 | 45.15 | 763 | 360 | 0.472 | 1.799 | 826 | 1311 | 1.59 | 13.09 |
| American Society for Microbiology (ASM) | 29 | 101 | 3.48 | 9.42 | 775 | 612 | 0.790 | 2.640 | 804 | 713 | 0.89 | 3.17 |
| Emerald Group Publishing | 72 | 207 | 2.88 | 9.42 | 714 | 198 | 0.277 | 0.806 | 786 | 405 | 0.52 | 3.03 |
| Psychology Press | 60 | 364 | 6.07 | 10.12 | 722 | 531 | 0.735 | 2.380 | 782 | 895 | 1.14 | 3.87 |
| Transaction Publishing | 82 | 83 | 1.01 | 1.62 | 632 | 41 | 0.065 | 0.411 | 714 | 124 | 0.17 | 0.73 |
| Baylor University Press | 70 | 39 | 0.56 | 1.00 | 625 | 4 | 0.006 | 0.080 | 695 | 43 | 0.06 | 0.36 |
| Studium Press | 38 | 7 | 0.18 | 0.69 | 639 | 112 | 0.175 | 0.566 | 677 | 119 | 0.18 | 0.57 |
| University of Washington Press | 71 | 171 | 2.41 | 5.18 | 590 | 17 | 0.029 | 0.195 | 661 | 188 | 0.28 | 1.85 |
| Policy Press (University of Bristol) | 57 | 70 | 1.23 | 3.96 | 602 | 77 | 0.128 | 0.558 | 659 | 147 | 0.22 | 1.31 |
| William Andrew | 41 | 149 | 3.63 | 5.98 | 605 | 37 | 0.061 | 0.332 | 646 | 186 | 0.29 | 1.76 |
| Pickering & Chatto | 77 | 88 | 1.14 | 1.33 | 533 | 2 | 0.004 | 0.061 | 610 | 90 | 0.15 | 0.61 |
| SIAM | 57 | 738 | 12.95 | 23.32 | 551 | 40 | 0.073 | 0.689 | 608 | 778 | 1.28 | 8.05 |
| IWA Publishing | 45 | 132 | 2.93 | 5.05 | 557 | 108 | 0.194 | 0.750 | 602 | 240 | 0.40 | 1.71 |
| Quintessence Publishing Co Inc | 36 | 23 | 0.64 | 1.17 | 474 | 76 | 0.160 | 0.551 | 510 | 99 | 0.19 | 0.63 |
| European Mathematical Society | 51 | 213 | 4.18 | 5.45 | 447 | 137 | 0.306 | 2.007 | 498 | 350 | 0.70 | 2.82 |
| Wilfrid Laurier University Press | 34 | 80 | 2.35 | 2.65 | 462 | 62 | 0.134 | 0.881 | 496 | 142 | 0.29 | 1.23 |
| Utah State University Press | 42 | 74 | 1.76 | 2.36 | 438 | 48 | 0.110 | 0.707 | 480 | 122 | 0.25 | 1.07 |
| Purdue University Press | 31 | 54 | 1.74 | 2.71 | 431 | 4 | 0.009 | 0.118 | 462 | 58 | 0.13 | 0.82 |
| Maney Publishing | 53 | 128 | 2.42 | 2.26 | 408 | 42 | 0.103 | 0.500 | 461 | 170 | 0.37 | 1.16 |
| Karger | 22 | 52 | 2.36 | 4.67 | 438 | 131 | 0.299 | 0.910 | 460 | 183 | 0.40 | 1.41 |
| Vanderbilt University Press | 37 | 60 | 1.62 | 2.05 | 406 | 15 | 0.037 | 0.246 | 443 | 75 | 0.17 | 0.77 |
| T.M.C. Asser Press | 35 | 65 | 1.86 | 2.64 | 406 | 92 | 0.227 | 0.725 | 441 | 157 | 0.36 | 1.10 |
| Johns Hopkins University Press | 66 | 149 | 2.26 | 2.37 | 333 | 6 | 0.018 | 0.154 | 399 | 155 | 0.39 | 1.28 |
| Wageningen | 32 | 53 | 1.66 | 2.15 | 357 | 86 | 0.241 | 0.656 | 389 | 139 | 0.36 | 0.96 |
| U.S. Institute of Peace | 39 | 181 | 4.64 | 9.43 | 327 | 92 | 0.281 | 0.978 | 366 | 273 | 0.75 | 3.45 |
| Academia Press | 25 | 16 | 0.64 | 1.08 | 316 | 2 | 0.006 | 0.079 | 341 | 18 | 0.05 | 0.34 |
| Sydney University Press | 23 | 47 | 2.04 | 2.40 | 316 | 22 | 0.070 | 0.331 | 339 | 69 | 0.20 | 0.85 |
| University of North Texas Press | 22 | 30 | 1.36 | 1.79 | 314 | 1 | 0.003 | 0.056 | 336 | 31 | 0.09 | 0.56 |
| WIT Press | 28 | 88 | 3.14 | 6.52 | 307 | 29 | 0.094 | 0.512 | 335 | 117 | 0.35 | 2.10 |
| World Health Organization | 45 | 3385 | 75.22 | 171.75 | 268 | 187 | 0.698 | 3.469 | 313 | 3572 | 11.41 | 69.69 |
| Institute of Southeast Asian Studies | 27 | 23 | 0.85 | 1.13 | 239 | 14 | 0.059 | 0.312 | 266 | 37 | 0.14 | 0.52 |
| ASTM International | 12 | 15 | 1.25 | 1.96 | 248 | 1 | 0.004 | 0.064 | 260 | 16 | 0.06 | 0.49 |
| University of Alaska Press | 17 | 77 | 4.53 | 9.15 | 226 | 9 | 0.040 | 0.255 | 243 | 86 | 0.35 | 2.63 |
| Nottingham University Press | 19 | 16 | 0.84 | 1.17 | 221 | 16 | 0.072 | 0.336 | 240 | 32 | 0.13 | 0.50 |
| SAE International | 16 | 13 | 0.81 | 1.17 | 200 | 3 | 0.015 | 0.122 | 216 | 16 | 0.07 | 0.39 |
| W. Bertelsmann Verlag GmbH | 18 | 8 | 0.44 | 0.92 | 196 | 5 | 0.026 | 0.158 | 214 | 13 | 0.06 | 0.32 |
| Athabasca University Press | 17 | 30 | 1.76 | 2.17 | 188 | 2 | 0.011 | 0.103 | 205 | 32 | 0.16 | 0.78 |
| Mary Ann Liebert | 6 | 15 | 2.50 | 5.17 | 197 | 73 | 0.371 | 0.869 | 203 | 88 | 0.43 | 1.23 |
| Ateneo de Manila University | 20 | 20 | 1.00 | 2.13 | 181 | 3 | 0.017 | 0.128 | 201 | 23 | 0.11 | 0.73 |
| ME Sharpe | 12 | 16 | 1.33 | 2.23 | 187 | 47 | 0.251 | 0.730 | 199 | 63 | 0.32 | 0.92 |
| American Psychiatric Publishing | 9 | 63 | 7.00 | 6.14 | 186 | 91 | 0.489 | 1.529 | 195 | 154 | 0.79 | 2.38 |
| University of Chicago Press | 18 | 37 | 2.06 | 1.63 | 166 | 7 | 0.042 | 0.472 | 184 | 44 | 0.24 | 0.90 |
| Anthem Press | 14 | 11 | 0.79 | 1.25 | 169 | 5 | 0.030 | 0.170 | 183 | 16 | 0.09 | 0.42 |
| University of Georgia Press | 17 | 29 | 1.71 | 4.07 | 161 | 0 | 0.000 | 0.000 | 178 | 29 | 0.16 | 1.32 |
| Universidad de Guadalajara | 18 | 0 | 0.00 | 0.00 | 159 | 0 | 0.000 | 0.000 | 177 | 0 | 0.00 | 0.00 |
| Spon Press | 14 | 59 | 4.21 | 8.64 | 144 | 5 | 0.035 | 0.275 | 158 | 64 | 0.41 | 2.77 |
| Editions Technip - Technical books | 11 | 13 | 1.18 | 1.83 | 145 | 15 | 0.103 | 0.574 | 156 | 28 | 0.18 | 0.77 |
| Ios Press | 10 | 14 | 1.40 | 2.72 | 145 | 22 | 0.152 | 0.730 | 155 | 36 | 0.23 | 1.01 |
| Physica-Verlag Gmbh & Co | 9 | 5 | 0.56 | 0.88 | 143 | 82 | 0.573 | 1.904 | 152 | 87 | 0.57 | 1.86 |
| VS Verlag für Sozialwissenschaften | 12 | 0 | 0.00 | 0.00 | 137 | 25 | 0.182 | 0.518 | 149 | 25 | 0.17 | 0.50 |
| Martinus Nijhoff | 10 | 0 | 0.00 | 0.00 | 137 | 2 | 0.015 | 0.120 | 147 | 2 | 0.01 | 0.12 |
| Brandeis University Press | 13 | 19 | 1.46 | 1.90 | 125 | 0 | 0.000 | 0.000 | 138 | 19 | 0.14 | 0.71 |
| Sinauer Associates | 10 | 100 | 10.00 | 8.76 | 123 | 0 | 0.000 | 0.000 | 133 | 100 | 0.75 | 3.50 |
| University of Adelaide Press | 9 | 6 | 0.67 | 1.32 | 124 | 0 | 0.000 | 0.000 | 133 | 6 | 0.05 | 0.37 |
| Dartmouth College Press | 11 | 9 | 0.82 | 0.98 | 120 | 1 | 0.008 | 0.091 | 131 | 10 | 0.08 | 0.36 |
| Science Press Beijing | 11 | 63 | 5.73 | 8.26 | 119 | 5 | 0.042 | 0.377 | 130 | 68 | 0.52 | 2.82 |
| Baywood Publishing | 11 | 7 | 0.64 | 0.92 | 112 | 29 | 0.259 | 0.681 | 123 | 36 | 0.29 | 0.71 |
| Geological Society UK | 7 | 7 | 1.00 | 1.00 | 115 | 332 | 2.887 | 4.397 | 122 | 339 | 2.78 | 4.30 |
| American Fisheries Society | 6 | 38 | 6.33 | 13.57 | 108 | 167 | 1.546 | 2.381 | 114 | 205 | 1.80 | 3.83 |
| American Society of Mechanical Engineers | 7 | 19 | 2.71 | 3.82 | 106 | 4 | 0.038 | 0.236 | 113 | 23 | 0.20 | 1.12 |
| Central European University Press | 8 | 25 | 3.13 | 2.47 | 94 | 4 | 0.043 | 0.203 | 102 | 29 | 0.28 | 1.08 |
| Northeastern University Press | 8 | 23 | 2.88 | 2.90 | 89 | 0 | 0.000 | 0.000 | 97 | 23 | 0.24 | 1.12 |
| Camel Publishing House | 1 | 0 | 0.00 | 0.00 | 92 | 0 | 0.000 | 0.000 | 93 | 0 | 0.00 | 0.00 |
| Dunedin Academic Press | 8 | 32 | 4.00 | 3.85 | 84 | 4 | 0.048 | 0.265 | 92 | 36 | 0.39 | 1.57 |
| Society for Mining, Metallurgy and Exploration | 4 | 7 | 1.75 | 3.50 | 86 | 23 | 0.267 | 0.622 | 90 | 30 | 0.33 | 0.94 |
| Universidad Nacional Autonoma de Mexico (UNAM) | 9 | 0 | 0.00 | 0.00 | 79 | 3 | 0.038 | 0.338 | 88 | 3 | 0.03 | 0.32 |
| DesTech | 5 | 1 | 0.20 | 0.45 | 81 | 3 | 0.037 | 0.190 | 86 | 4 | 0.05 | 0.21 |
| John Benjamins | 10 | 35 | 3.50 | 3.06 | 76 | 7 | 0.092 | 0.334 | 86 | 42 | 0.49 | 1.52 |
| Minerva Medica | 1 | 3 | 3.00 | 0.00 | 84 | 5 | 0.060 | 0.284 | 85 | 8 | 0.09 | 0.43 |
| ISTE Ltd | 8 | 8 | 1.00 | 2.83 | 76 | 4 | 0.053 | 0.361 | 84 | 12 | 0.14 | 0.93 |
| No identificada | 4 | 2 | 0.50 | 0.58 | 80 | 38 | 0.475 | 1.273 | 84 | 40 | 0.48 | 1.25 |
| Palacký University | 9 | 0 | 0.00 | 0.00 | 75 | 4 | 0.053 | 0.280 | 84 | 4 | 0.05 | 0.26 |
| Univerzita Palackého v Olomouci | 8 | 0 | 0.00 | 0.00 | 76 | 0 | 0.000 | 0.000 | 84 | 0 | 0.00 | 0.00 |
| Mathematical Association of America | 4 | 1 | 0.25 | 0.50 | 77 | 0 | 0.000 | 0.000 | 81 | 1 | 0.01 | 0.11 |
| NSTA Press - National Science Teachers Association | 7 | 7 | 1.00 | 1.91 | 72 | 11 | 0.153 | 0.465 | 79 | 18 | 0.23 | 0.73 |
| American Institute of Aeronautics and Astronautics | 5 | 0 | 0.00 | 0.00 | 73 | 18 | 0.247 | 1.762 | 78 | 18 | 0.23 | 1.71 |
| EAGE (European Association of Geoscientists & Engineers | 6 | 11 | 1.83 | 1.47 | 65 | 0 | 0.000 | 0.000 | 71 | 11 | 0.15 | 0.65 |
| International Monetary Fund (IMF) | 5 | 3 | 0.60 | 0.55 | 66 | 1 | 0.015 | 0.123 | 71 | 4 | 0.06 | 0.23 |
| Ernst & Sohn | 4 | 2 | 0.50 | 0.58 | 64 | 1 | 0.016 | 0.125 | 68 | 3 | 0.04 | 0.21 |
| Austrian Academy of Sciences | 3 | 7 | 2.33 | 4.04 | 59 | 0 | 0.000 | 0.000 | 62 | 7 | 0.11 | 0.89 |
| University of New Hampshire Press | 6 | 10 | 1.67 | 1.97 | 55 | 7 | 0.127 | 0.610 | 61 | 17 | 0.28 | 0.93 |
| White Horse Press | 6 | 19 | 3.17 | 2.93 | 53 | 10 | 0.189 | 0.761 | 59 | 29 | 0.49 | 1.44 |
| Imperial College Press | 4 | 3 | 0.75 | 1.50 | 52 | 5 | 0.096 | 0.298 | 56 | 8 | 0.14 | 0.48 |
| InTech | 3 | 3 | 1.00 | 1.73 | 53 | 11 | 0.208 | 0.689 | 56 | 14 | 0.25 | 0.77 |
| Rodopi | 3 | 7 | 2.33 | 3.21 | 52 | 4 | 0.077 | 0.334 | 55 | 11 | 0.20 | 0.87 |
| Tsinghua University | 4 | 13 | 3.25 | 2.50 | 51 | 23 | 0.451 | 1.331 | 55 | 36 | 0.65 | 1.59 |
| University of Alberta | 3 | 20 | 6.67 | 9.07 | 48 | 64 | 1.333 | 2.263 | 51 | 84 | 1.65 | 3.12 |
| Equinox | 4 | 17 | 4.25 | 5.97 | 44 | 9 | 0.205 | 0.795 | 48 | 26 | 0.54 | 2.03 |
| Kluwer Academic Publishers | 3 | 66 | 22.00 | 18.19 | 45 | 41 | 0.911 | 1.844 | 48 | 107 | 2.23 | 6.62 |
| IOP Publishing | 1 | 3 | 3.00 | 0.00 | 46 | 5 | 0.109 | 0.315 | 47 | 8 | 0.17 | 0.52 |
| Monash University Press | 3 | 16 | 5.33 | 2.08 | 44 | 8 | 0.182 | 0.582 | 47 | 24 | 0.51 | 1.46 |
| Martin Striz | 3 | 0 | 0.00 | 0.00 | 41 | 2 | 0.049 | 0.218 | 44 | 2 | 0.05 | 0.21 |
| Pabst Science Publishers | 3 | 14 | 4.67 | 4.04 | 41 | 35 | 0.854 | 1.442 | 44 | 49 | 1.11 | 1.91 |
| Univerza v Ljubljani | 6 | 0 | 0.00 | 0.00 | 38 | 1 | 0.026 | 0.162 | 44 | 1 | 0.02 | 0.15 |
| Praxis Publishing Ltd. | 3 | 5 | 1.67 | 2.08 | 40 | 0 | 0.000 | 0.000 | 43 | 5 | 0.12 | 0.63 |
| Earthscan Publications Ltd. | 4 | 5 | 1.25 | 1.89 | 37 | 1 | 0.027 | 0.164 | 41 | 6 | 0.15 | 0.65 |
| IEEE | 3 | 0 | 0.00 | 0.00 | 38 | 3 | 0.079 | 0.273 | 41 | 3 | 0.07 | 0.26 |
| Royal College of Psychiatrists | 3 | 16 | 5.33 | 6.81 | 37 | 16 | 0.432 | 1.324 | 40 | 32 | 0.80 | 2.39 |
| American Geophysical Union | 3 | 5 | 1.67 | 1.15 | 36 | 51 | 1.417 | 2.970 | 39 | 56 | 1.44 | 2.86 |
| Syngress | 4 | 0 | 0.00 | 0.00 | 34 | 0 | 0.000 | 0.000 | 38 | 0 | 0.00 | 0.00 |
| Resources for the Future | 2 | 14 | 7.00 | 4.24 | 34 | 19 | 0.559 | 0.927 | 36 | 33 | 0.92 | 1.89 |
| ARG Gantner VerlaG | 8 | 13 | 1.63 | 1.77 | 27 | 13 | 0.481 | 1.221 | 35 | 26 | 0.74 | 1.42 |
| Instytut Badawczy Leśnictwa: | 3 | 0 | 0.00 | 0.00 | 31 | 0 | 0.000 | 0.000 | 34 | 0 | 0.00 | 0.00 |
| Oxford University Press | 3 | 33 | 11.00 | 14.80 | 31 | 0 | 0.000 | 0.000 | 34 | 33 | 0.97 | 4.83 |
| World Scientific and Engineering Academy and Society | 2 | 4 | 2.00 | 2.83 | 32 | 8 | 0.250 | 0.508 | 34 | 12 | 0.35 | 0.81 |
| Institute of Systematics and Evolution of Animals | 7 | 14 | 2.00 | 4.12 | 24 | 12 | 0.500 | 1.063 | 31 | 26 | 0.84 | 2.16 |
| Newnes | 2 | 1 | 0.50 | 0.71 | 29 | 0 | 0.000 | 0.000 | 31 | 1 | 0.03 | 0.18 |
| Cold Spring Harbor Press | 3 | 1 | 0.33 | 0.58 | 27 | 23 | 0.852 | 0.949 | 30 | 24 | 0.80 | 0.92 |
| Anderson Publishing | 1 | 153 | 153.00 | 0.00 | 28 | 1 | 0.036 | 0.189 | 29 | 154 | 5.31 | 28.41 |
| Monduzzi Editoriale | 1 | 0 | 0.00 | 0.00 | 28 | 0 | 0.000 | 0.000 | 29 | 0 | 0.00 | 0.00 |
| Oxford Fajar Sdn. Bhd. | 1 | 1 | 1.00 | 0.00 | 28 | 4 | 0.143 | 0.448 | 29 | 5 | 0.17 | 0.47 |
| Society for the Promotion of Science and Scholarship | 1 | 0 | 0.00 | 0.00 | 27 | 0 | 0.000 | 0.000 | 28 | 0 | 0.00 | 0.00 |
| Darlington Press | 3 | 13 | 4.33 | 3.06 | 24 | 0 | 0.000 | 0.000 | 27 | 13 | 0.48 | 1.63 |
| Fim de Século Edições | 2 | 0 | 0.00 | 0.00 | 24 | 0 | 0.000 | 0.000 | 26 | 0 | 0.00 | 0.00 |
| Metropolitan University Prague | 1 | 0 | 0.00 | 0.00 | 25 | 0 | 0.000 | 0.000 | 26 | 0 | 0.00 | 0.00 |
| University Press of New England | 2 | 3 | 1.50 | 0.71 | 24 | 0 | 0.000 | 0.000 | 26 | 3 | 0.12 | 0.43 |
| EAC - Europae Archaeologiae Consilium | 1 | 4 | 4.00 | 0.00 | 24 | 1 | 0.042 | 0.204 | 25 | 5 | 0.20 | 0.82 |
| Lars Müller Publishers | 3 | 0 | 0.00 | 0.00 | 21 | 1 | 0.048 | 0.218 | 24 | 1 | 0.04 | 0.20 |
| Duke University Press | 1 | 0 | 0.00 | 0.00 | 22 | 4 | 0.182 | 0.664 | 23 | 4 | 0.17 | 0.65 |
| Fisheries and Oceans Canada | 1 | 0 | 0.00 | 0.00 | 22 | 0 | 0.000 | 0.000 | 23 | 0 | 0.00 | 0.00 |
| Gallaudet University Press | 2 | 3 | 1.50 | 0.71 | 20 | 1 | 0.050 | 0.224 | 22 | 4 | 0.18 | 0.50 |
| Trans Tech Publications Ltd | 3 | 1 | 0.33 | 0.58 | 19 | 4 | 0.211 | 0.713 | 22 | 5 | 0.23 | 0.69 |
| University of Tartu Press | 1 | 0 | 0.00 | 0.00 | 21 | 0 | 0.000 | 0.000 | 22 | 0 | 0.00 | 0.00 |
| ChemTec Publishing | 1 | 17 | 17.00 | 0.00 | 20 | 0 | 0.000 | 0.000 | 21 | 17 | 0.81 | 3.71 |
| Jones & Bartlett Learning | 1 | 1 | 1.00 | 0.00 | 20 | 3 | 0.150 | 0.489 | 21 | 4 | 0.19 | 0.51 |
| Siam Society Under Royal Patronage | 1 | 0 | 0.00 | 0.00 | 20 | 0 | 0.000 | 0.000 | 21 | 0 | 0.00 | 0.00 |
| ASM International | 1 | 2 | 2.00 | 0.00 | 19 | 0 | 0.000 | 0.000 | 20 | 2 | 0.10 | 0.45 |
| North-European Scientific Publishers | 1 | 0 | 0.00 | 0.00 | 19 | 0 | 0.000 | 0.000 | 20 | 0 | 0.00 | 0.00 |
| The Forest History Society | 2 | 5 | 2.50 | 3.54 | 18 | 1 | 0.056 | 0.236 | 20 | 6 | 0.30 | 1.13 |
| Vilnius Gediminas Technical University Press | 3 | 0 | 0.00 | 0.00 | 17 | 0 | 0.000 | 0.000 | 20 | 0 | 0.00 | 0.00 |
| Asclepion Publishing | 1 | 4 | 4.00 | 0.00 | 18 | 0 | 0.000 | 0.000 | 19 | 4 | 0.21 | 0.92 |
| Chartered Institute of Management Accountants | 3 | 3 | 1.00 | 1.73 | 16 | 0 | 0.000 | 0.000 | 19 | 3 | 0.16 | 0.69 |
| New Zealand Freshwater Sciences Society | 1 | 1 | 1.00 | 0.00 | 18 | 0 | 0.000 | 0.000 | 19 | 1 | 0.05 | 0.23 |
| ASERS Publishing | 2 | 0 | 0.00 | 0.00 | 16 | 0 | 0.000 | 0.000 | 18 | 0 | 0.00 | 0.00 |
| Corila | 1 | 5 | 5.00 | 0.00 | 17 | 19 | 1.118 | 1.965 | 18 | 24 | 1.33 | 2.11 |
| Scuola Normale Superiore di Pisa | 2 | 0 | 0.00 | 0.00 | 16 | 7 | 0.438 | 1.504 | 18 | 7 | 0.39 | 1.42 |
| Slovenska Akademija Znanosti in Umetnosti | 1 | 0 | 0.00 | 0.00 | 17 | 1 | 0.059 | 0.243 | 18 | 1 | 0.06 | 0.24 |
| University of Wales Press | 1 | 0 | 0.00 | 0.00 | 17 | 0 | 0.000 | 0.000 | 18 | 0 | 0.00 | 0.00 |
| Cooper Ornithological Society | 2 | 2 | 1.00 | 1.41 | 15 | 1 | 0.067 | 0.258 | 17 | 3 | 0.18 | 0.53 |
| Museu Nacional - UFRJ - | 2 | 0 | 0.00 | 0.00 | 15 | 17 | 1.133 | 1.767 | 17 | 17 | 1.00 | 1.70 |
| South African National Biodiversity Institute | 2 | 6 | 3.00 | 1.41 | 15 | 3 | 0.200 | 0.561 | 17 | 9 | 0.53 | 1.12 |
| Indian National Science Academy | 1 | 0 | 0.00 | 0.00 | 15 | 10 | 0.667 | 1.839 | 16 | 10 | 0.63 | 1.78 |
| Island Press | 1 | 1 | 1.00 | 0.00 | 15 | 4 | 0.267 | 0.594 | 16 | 5 | 0.31 | 0.60 |
| Lawrence Erlbaum Associates | 1 | 3 | 3.00 | 0.00 | 15 | 12 | 0.800 | 0.941 | 16 | 15 | 0.94 | 1.06 |
| Scrivener Publishing | 1 | 0 | 0.00 | 0.00 | 15 | 0 | 0.000 | 0.000 | 16 | 0 | 0.00 | 0.00 |
| Global Acadamic Publishing Binghamton | 1 | 0 | 0.00 | 0.00 | 14 | 0 | 0.000 | 0.000 | 15 | 0 | 0.00 | 0.00 |
| Verlag Volk und Gesundheit | 1 | 27 | 27.00 | 0.00 | 14 | 14 | 1.000 | 3.211 | 15 | 41 | 2.73 | 7.39 |
| Yale University Press | 1 | 0 | 0.00 | 0.00 | 14 | 11 | 0.786 | 2.940 | 15 | 11 | 0.73 | 2.84 |
| Academic Conferences Limited | 1 | 1 | 1.00 | 0.00 | 13 | 0 | 0.000 | 0.000 | 14 | 1 | 0.07 | 0.27 |
| Academy of Sciences of the Czech Republic - | 1 | 0 | 0.00 | 0.00 | 13 | 2 | 0.154 | 0.555 | 14 | 2 | 0.14 | 0.53 |
| Burke Museum of Natural History and Culture | 1 | 0 | 0.00 | 0.00 | 13 | 1 | 0.077 | 0.277 | 14 | 1 | 0.07 | 0.27 |
| Channel View Publications | 1 | 3 | 3.00 | 0.00 | 13 | 0 | 0.000 | 0.000 | 14 | 3 | 0.21 | 0.80 |
| Fouqué Literaturverlag | 1 | 0 | 0.00 | 0.00 | 13 | 0 | 0.000 | 0.000 | 14 | 0 | 0.00 | 0.00 |
| Modern Humanities Research Association | 4 | 6 | 1.50 | 1.29 | 10 | 0 | 0.000 | 0.000 | 14 | 6 | 0.43 | 0.94 |
| Royal Historical Society (RHS) | 5 | 3 | 0.60 | 1.34 | 9 | 0 | 0.000 | 0.000 | 14 | 3 | 0.21 | 0.80 |
| Vydavatelství a nakladatelství Aleš Čeněk | 1 | 0 | 0.00 | 0.00 | 13 | 0 | 0.000 | 0.000 | 14 | 0 | 0.00 | 0.00 |
| Analytics Press | 3 | 3 | 1.00 | 1.00 | 10 | 0 | 0.000 | 0.000 | 13 | 3 | 0.23 | 0.60 |
| Australasian Institute of Mining and Metallurgy | 1 | 0 | 0.00 | 0.00 | 12 | 10 | 0.833 | 1.337 | 13 | 10 | 0.77 | 1.30 |
| CA Press - CA Technologies | 1 | 0 | 0.00 | 0.00 | 12 | 0 | 0.000 | 0.000 | 13 | 0 | 0.00 | 0.00 |
| Éditions L'Harmattan | 1 | 0 | 0.00 | 0.00 | 12 | 0 | 0.000 | 0.000 | 13 | 0 | 0.00 | 0.00 |
| Freund Publishing House Ltd | 1 | 0 | 0.00 | 0.00 | 12 | 0 | 0.000 | 0.000 | 13 | 0 | 0.00 | 0.00 |
| Kogan Page | 1 | 0 | 0.00 | 0.00 | 12 | 0 | 0.000 | 0.000 | 13 | 0 | 0.00 | 0.00 |
| Productivity Press Publishing | 1 | 3 | 3.00 | 0.00 | 12 | 0 | 0.000 | 0.000 | 13 | 3 | 0.23 | 0.83 |
| Embrapa Soja | 1 | 0 | 0.00 | 0.00 | 11 | 2 | 0.182 | 0.405 | 12 | 2 | 0.17 | 0.39 |
| Poly Ppress | 1 | 0 | 0.00 | 0.00 | 11 | 0 | 0.000 | 0.000 | 12 | 0 | 0.00 | 0.00 |
| Professional & Higher (UK) | 1 | 1 | 1.00 | 0.00 | 11 | 0 | 0.000 | 0.000 | 12 | 1 | 0.08 | 0.29 |
| Prometheus Books | 3 | 1 | 0.33 | 0.58 | 9 | 0 | 0.000 | 0.000 | 12 | 1 | 0.08 | 0.29 |
| ProQuest | 2 | 0 | 0.00 | 0.00 | 10 | 0 | 0.000 | 0.000 | 12 | 0 | 0.00 | 0.00 |
| Science History Publications | 1 | 0 | 0.00 | 0.00 | 11 | 53 | 4.818 | 13.688 | 12 | 53 | 4.42 | 13.12 |
| University of New Mexico Press | 1 | 0 | 0.00 | 0.00 | 11 | 0 | 0.000 | 0.000 | 12 | 0 | 0.00 | 0.00 |
| AICA-Armenia | 1 | 0 | 0.00 | 0.00 | 10 | 0 | 0.000 | 0.000 | 11 | 0 | 0.00 | 0.00 |
| Asociación Española de Micología | 1 | 0 | 0.00 | 0.00 | 10 | 0 | 0.000 | 0.000 | 11 | 0 | 0.00 | 0.00 |
| Cambridge International Science Publishing Ltd. | 1 | 0 | 0.00 | 0.00 | 10 | 0 | 0.000 | 0.000 | 11 | 0 | 0.00 | 0.00 |
| Deutsches Grünes Kreuz e.V. | 1 | 0 | 0.00 | 0.00 | 10 | 0 | 0.000 | 0.000 | 11 | 0 | 0.00 | 0.00 |
| Focal Press | 1 | 0 | 0.00 | 0.00 | 10 | 0 | 0.000 | 0.000 | 11 | 0 | 0.00 | 0.00 |
| Mohr Siebeck Verlag | 1 | 7 | 7.00 | 0.00 | 10 | 0 | 0.000 | 0.000 | 11 | 7 | 0.64 | 2.11 |
| Saint Joseph's University Press | 1 | 1 | 1.00 | 0.00 | 10 | 0 | 0.000 | 0.000 | 11 | 1 | 0.09 | 0.30 |
| University of West Bohemia | 1 | 0 | 0.00 | 0.00 | 10 | 0 | 0.000 | 0.000 | 11 | 0 | 0.00 | 0.00 |
| British Psychological Society | 1 | 3 | 3.00 | 0.00 | 9 | 0 | 0.000 | 0.000 | 10 | 3 | 0.30 | 0.95 |
| Edizioni della Scuola Normale Superiore | 3 | 4 | 1.33 | 1.53 | 7 | 3 | 0.429 | 1.134 | 10 | 7 | 0.70 | 1.25 |
| Lincoln Institute of Land Policy | 1 | 7 | 7.00 | 0.00 | 9 | 0 | 0.000 | 0.000 | 10 | 7 | 0.70 | 2.21 |
| Lynx House Press | 1 | 0 | 0.00 | 0.00 | 9 | 0 | 0.000 | 0.000 | 10 | 0 | 0.00 | 0.00 |
| Proverse Hong Kong | 1 | 0 | 0.00 | 0.00 | 9 | 0 | 0.000 | 0.000 | 10 | 0 | 0.00 | 0.00 |
| Associação Brasileira de Mecânica dos Solos e Engenharia Geotécnica | 1 | 0 | 0.00 | 0.00 | 8 | 0 | 0.000 | 0.000 | 9 | 0 | 0.00 | 0.00 |
| Cesi - Casa Editrice Scientifica Internazionale | 1 | 0 | 0.00 | 0.00 | 8 | 0 | 0.000 | 0.000 | 9 | 0 | 0.00 | 0.00 |
| InterVarsity Press | 1 | 0 | 0.00 | 0.00 | 8 | 0 | 0.000 | 0.000 | 9 | 0 | 0.00 | 0.00 |
| Koeltz Scientific Books | 2 | 2 | 1.00 | 1.41 | 7 | 0 | 0.000 | 0.000 | 9 | 2 | 0.22 | 0.67 |
| Research Institute on Mediterranean Civilizations | 1 | 0 | 0.00 | 0.00 | 8 | 0 | 0.000 | 0.000 | 9 | 0 | 0.00 | 0.00 |
| Society for Nautical Research | 1 | 4 | 4.00 | 0.00 | 8 | 0 | 0.000 | 0.000 | 9 | 4 | 0.44 | 1.33 |
| East-West Center | 1 | 9 | 9.00 | 0.00 | 7 | 0 | 0.000 | 0.000 | 8 | 9 | 1.13 | 3.18 |
| École polytechnique | 1 | 0 | 0.00 | 0.00 | 7 | 0 | 0.000 | 0.000 | 8 | 0 | 0.00 | 0.00 |
| S. P. Timoshenko Institute of Mechanics | 1 | 0 | 0.00 | 0.00 | 7 | 0 | 0.000 | 0.000 | 8 | 0 | 0.00 | 0.00 |
| Silkworm Books | 1 | 0 | 0.00 | 0.00 | 7 | 0 | 0.000 | 0.000 | 8 | 0 | 0.00 | 0.00 |
| USDA Forest Service | 1 | 0 | 0.00 | 0.00 | 7 | 1 | 0.143 | 0.378 | 8 | 1 | 0.13 | 0.35 |
| Wellcome Trust Centre for the History of Medicine | 1 | 0 | 0.00 | 0.00 | 7 | 3 | 0.429 | 0.535 | 8 | 3 | 0.38 | 0.52 |
| Editura CuArt | 3 | 0 | 0.00 | 0.00 | 4 | 0 | 0.000 | 0.000 | 7 | 0 | 0.00 | 0.00 |
| Infonortics Ltd | 1 | 1 | 1.00 | 0.00 | 6 | 0 | 0.000 | 0.000 | 7 | 1 | 0.14 | 0.38 |
| Slovak Academy of Sciences | 1 | 0 | 0.00 | 0.00 | 6 | 3 | 0.500 | 0.548 | 7 | 3 | 0.43 | 0.53 |
| Zip Publishing | 1 | 0 | 0.00 | 0.00 | 6 | 0 | 0.000 | 0.000 | 7 | 0 | 0.00 | 0.00 |
| Instituto de Botânica - Governo do Estado de São Paulo | 2 | 0 | 0.00 | 0.00 | 4 | 0 | 0.000 | 0.000 | 6 | 0 | 0.00 | 0.00 |
| Mountain Press Publishing | 1 | 0 | 0.00 | 0.00 | 5 | 0 | 0.000 | 0.000 | 6 | 0 | 0.00 | 0.00 |
| Westdeutscher Verlag | 1 | 0 | 0.00 | 0.00 | 5 | 0 | 0.000 | 0.000 | 6 | 0 | 0.00 | 0.00 |
| Information Today, Inc | 1 | 0 | 0.00 | 0.00 | 3 | 0 | 0.000 | 0.000 | 4 | 0 | 0.00 | 0.00 |
| Sin Información | 1 | 0 | 0.00 | 0.00 | 3 | 0 | 0.000 | 0.000 | 4 | 0 | 0.00 | 0.00 |
| Catena Verlag | 1 | 0 | 0.00 | 0.00 | 1 | 5 | 5.000 | 0.00 | 2 | 5 | 2.50 | 3.54 |
| Conchbooks | 1 | 0 | 0.00 | 0.00 | 1 | 0 | 0.000 | 0.00 | 2 | 0 | 0.00 | 0.00 |
| Muzeum i Instytut Zoologii | 1 | 0 | 0.00 | 0.00 | 1 | 5 | 5.000 | 0.00 | 2 | 5 | 2.50 | 3.54 |